\documentclass[pre,graphicx,twocolumn,superscriptaddress,unsortaddress,reprint,showpacs]{revtex4-1}

\usepackage{amsmath}
\usepackage{enumerate}
\usepackage{graphicx}
\usepackage{mathbbol}
\usepackage{amsfonts}
\usepackage{natbib}

\usepackage{bm}

\usepackage{color}

\newcommand{\rmd}{{\mathrm{d}}}

\begin{document}

\title{Maximum relative height of elastic interfaces in random media} 

\author{Joachim Rambeau}
\email[]{joachim.rambeau@th.u-psud.fr}
\affiliation{Laboratoire de Physique Th\'eorique d'Orsay, Universit\'e Paris Sud 11 and CNRS}

\author{Sebastian Bustingorry}
\email[]{sbusting@cab.cnea.gov.ar}
\affiliation{CONICET, Centro At\'omico Bariloche, 8400 San Carlos de Bariloche, R\'io Negro, Argentina}

\author{Alejandro B. Kolton}
\email[]{koltona@cab.cnea.gov.ar}
\affiliation{CONICET, Centro At\'omico Bariloche, 8400 San Carlos de Bariloche, R\'io Negro, Argentina}

\author{Gr\'egory Schehr}
\email[]{gregory.schehr@th.u-psud.fr}
\affiliation{Laboratoire de Physique Th\'eorique d'Orsay, Universit\'e Paris Sud 11 and CNRS}

\date{\today}

\begin{abstract}
The 
distribution of the maximal relative height (MRH) 
of self-affine one-dimensional elastic interfaces in a random 
potential is studied. We analyze the ground state configuration at zero driving force, 
and the critical configuration exactly at the 
depinning threshold, both for the random-manifold and random-periodic universality classes. 
These configurations are sampled by exact numerical methods, and  
their MRH distributions are compared with those with the same roughness exponent 
and boundary conditions, but produced by independent
Fourier modes with normally distributed amplitudes.  
Using Pickands' theorem we derive an exact analytical description for the right tail of 
the latter. After properly rescaling the MRH distributions we find that  
corrections from the Gaussian independent modes approximation
are in general small, as previously found for the average width distribution of depinning 
configurations. In the large size limit all corrections are finite except for the ground-state 
in the random-periodic class whose MRH distribution
becomes, for periodic boundary conditions, indistinguishable from the Airy distribution.
We find that the MRH distributions are, in general, sensitive 
to changes of boundary conditions.
\end{abstract}


\maketitle 

\section{Introduction}

Recent theoretical progress have highlighted the importance of extreme value
statistics (EVS) in statistical physics \cite{bouchaud_mezard,
bouchaud_biroli_review}. This often yields interesting questions of EVS for
strongly correlated variables, a field 
which, to a large extent, remains to be explored. For homogeneous systems, it has been
recently shown that
EVS is an interesting tool to characterize the statistics of one-dimensional
stochastic processes, like Brownian motion and its variants, which can be 
mapped onto models of elastic interfaces \cite{RCPS,satya_airyshort,
satya_airylong,schehr_airy,GMOR, RS09, RS10}.

Such elastic interfaces are ubiquitous in nature and in the simplest
one-dimensional case they can be 
parameterized by a scalar displacement field $u(x)$, for a system of length $L$.
Here we consider 
periodic boundary conditions (pbc), {\it i.e.} $u(x=0) = u(x=L)$ and focus on
the displacement, or relative {\it height}, 
$h(x)=u(x)- (1/L) \int_0^L \rmd x'\ u(x')$.
It can be decomposed into Fourier modes  
\begin{eqnarray}\label{exp_Fourier}
h(x) = \sum_{n=1}^\infty a_n \cos{\left(\frac{2 \pi n}{L} x \right)} + b_n
\sin{\left(\frac{2 \pi n}{L} x \right)} \;,
\end{eqnarray} 
where $a_n$'s and $b_n$'s are random variables, whose statistics depend on the
precise model under considerations. 
The geometry of such interfaces (\ref{exp_Fourier}) is usually characterized by
the sample-dependent roughness $w^2$ and its ensemble average 
${\rm w}_L^2$~\cite{barabasi_stanley}, 
\begin{equation}
w^2=(1/L) \int_0^L h(x)^2 \;, {\rm w}_L^2 = \langle w^2 \rangle, 
\end{equation}
where the brackets $\langle...\rangle$ denotes an average over $a_n$'s and
$b_n$'s. Of particular interest are self-affine (or critical) interfaces,  
characterized by a roughness exponent $\zeta$, such that ${\rm w}_L \sim
L^{\zeta}$. Relatively recently, the full distribution 
of $w^2$, not only the first moment ${\rm w}^2_L$, has been considered~\cite{width_gaussian}. It was indeed computed 
both for Gaussian interfaces, {\it i.e.} in the case
where $a_n$'s and $b_n$'s in Eq.~(\ref{exp_Fourier}) are independent Gaussian
variables of zero mean and variances $\langle a_n^2 \rangle = \langle b_n^2
\rangle \propto n^{-\alpha}$ [as in Eq. (\ref{Gauss}) below], corresponding to a
roughness exponent $\zeta = (\alpha-1)/2$, for $\alpha > 1$
\cite{width_gaussian}, as well as for pinned interfaces at the depinning
threshold \cite{width_rosso, width_frg}, where it was studied both numerically
and analytically using Functional Renormalization Group (FRG) \cite{review_frg}.
It was subsequently measured for a contact line in a wetting experiment, on a
disordered substrate \cite{moulinet_width}, and also generalized to Gaussian
``non-Markovian'' ($\alpha \neq 2$) stochastic processes~\cite{raoul_width}. 
It is interesting to note that such Gaussian interfaces can be physically regarded as 
an ensemble of free interfaces with non-local harmonic elastic interactions at 
thermal equilibrium, whose Hamiltonian can be elegantly written in terms of fractional 
derivatives of order $\alpha/2$~\cite{raoul_width}, 
\begin{equation}
 H[\{h(x)\}] = \frac{1}{2} \int_0^L \rmd x\;\left[\frac{d^{\alpha/2} h(x)}{dx^{\alpha/2}}\right]^2.
\label{eq:fractionalH}
\end{equation}
Here we focus on the maximal relative height (MRH) $h_{m}$ of such a fluctuating
interface (\ref{exp_Fourier}) defined as
\begin{equation}
\label{eq:MRH}
h_m=\max_{0\leq x \leq L} h(x) \;.
\end{equation}
This observable was first introduced and studied numerically for Gaussian
interfaces with short range elasticity, such that $\langle a_n^2 \rangle =
\langle b_n^2 \rangle \propto n^{-2}$ \cite{RCPS}, {\it i.e.} $\zeta = 1/2$, and
it was found that $\langle h_{m} \rangle \sim L^{\zeta}$. Then, Majumdar and Comtet
obtained an exact analytical expression for the full probability distribution
function (pdf) $P(h_m|L)$ of $h_m$ and showed that it is given by the so-called
Airy distribution $f_{\rm Airy}$
 ~\cite{satya_airyshort, satya_airylong},
\begin{eqnarray}\label{scaling_airy}
&& P(h_m|L)= \frac{1}{L^{1/2}}\, f_{\rm Airy}\left(\frac{h_m}{L^{1/2}}\right)
\;,
\end{eqnarray}
where $f_{\rm Airy}$ describes  the distribution of the area under a Brownian
excursion on the unit time interval. We remind that a Brownian
excursion is a Brownian motion conditioned to start and end in $0$ while staying
positive in the time interval $[0,1]$.

Incidentally, the same Airy distribution appears also in various, a priori unrelated
problems, in graph theory and 
in computer science \cite{majumdar_review}. This result (\ref{scaling_airy}) was
then extended to a wide class of one-dimensional solid-on-solid models
\cite{schehr_airy}, showing the universality of the Airy distribution
(\ref{scaling_airy}) for interface models with short range elasticity (and
without disorder). The distribution of $h_m$ was then investigated for Gaussian
interfaces with $\alpha > 1$ in Ref.~\cite{GMOR}. There it was shown that the
distribution has a scaling form similar to Eq. (\ref{scaling_airy}), the scaling
variable being $h_m/L^{\zeta}$, with $\zeta = (\alpha - 1)/2$, and where the scaling function depends
continuously on the parameter~$\alpha$.

\begin{figure}
\begin{center}
\includegraphics[width=.6\linewidth]{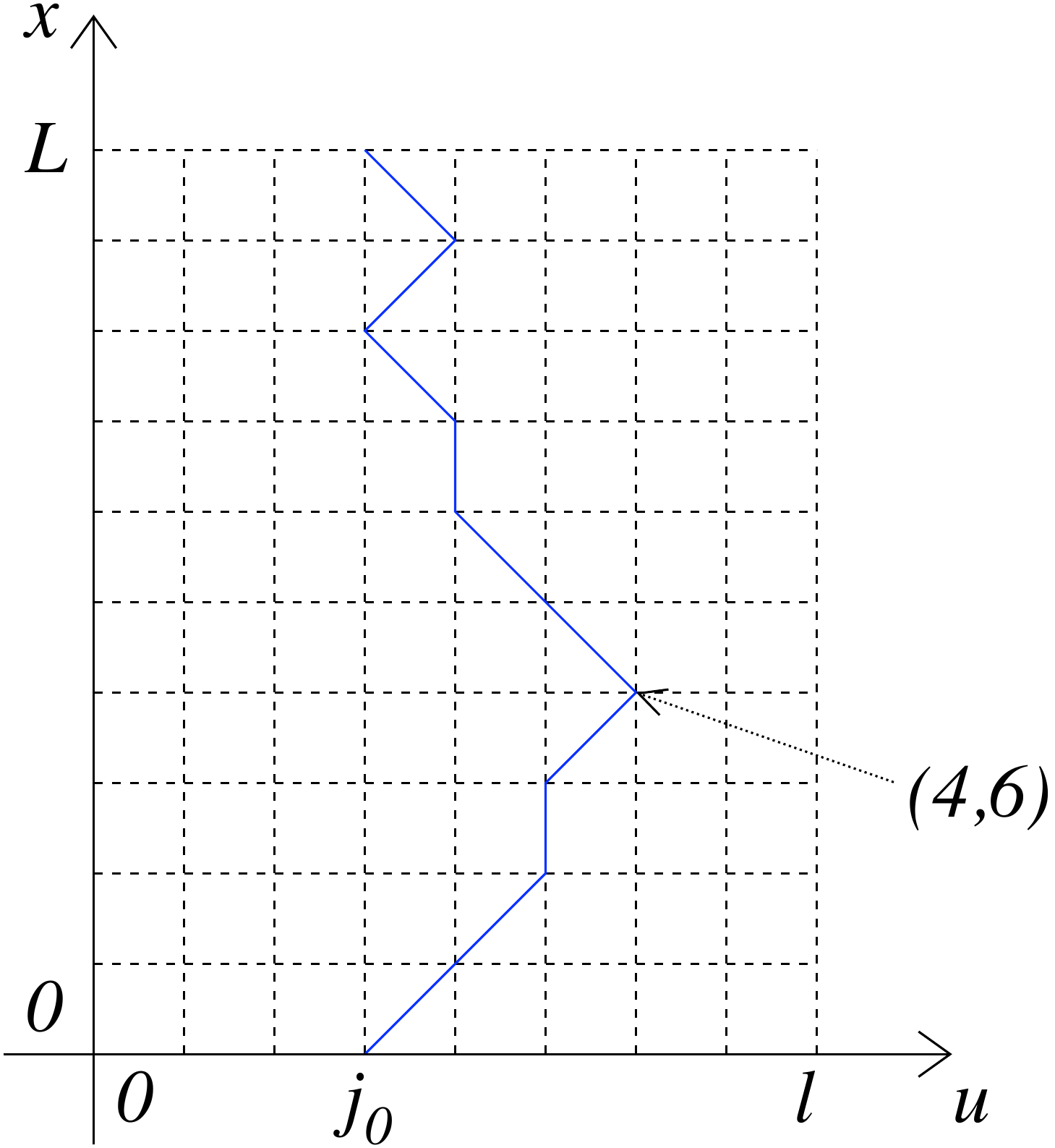}
\end{center}
\caption{The plain blue line represent a path on the lattice when starting in $j_0=3$.
The pointed maximum is reached at $i=4$ and its value is $j_4=6$. If $(j_i)_{0\leq i \leq L}$ is the sequence
of heights visited by the path, one must substract the spatial average $\langle u \rangle = \sum_{i} j_i /L$
to obtain the MRH of this sample.}
\label{fig_lattice}
\end{figure}

A natural, and very physical, extension of these works concerns the maximal
height of elastic interfaces {\it in random media}: this is the aim of the present
paper. In the continuum limit the simplest elastic interface model 
is described by the Hamiltonian
\begin{equation}
\label{energy}
H[\{u(x)\}]=\int_0^L \rmd x \left\{ \frac{c}{2} \left( \nabla u (x) \right)^2 +
V(x,u(x)) - f u(x)\right\} ,
\end{equation}
together with an overdamped equation of motion $\partial_t u(x,t) \propto -\delta H[\{u(y)\}]/\delta u(x)$.
The first term in~(\ref{energy}) is the usual harmonic elastic energy, where $c$ is the elastic constant,
and tends to straighten the displacement field. The second term is a random potential $V(x,u)$ that models 
randomly distributed defects both in the
direction $x$ along the line 
and in the direction $u$ along the displacements (see Fig. \ref{fig_lattice} for a discretized
version of this interface to be considered below). The third term represents 
the action of an external uniform field with strength 
$f$ that tends to drive the interface in direction $u$.
Here we consider a random
potential which is a Gaussian random variable with zero mean
$\overline{V(x,u)}=0$
and correlations
\begin{equation}
\label{eq:potential}
\overline{V(x,u) V(x',u')}= \delta(x-x') R_0(u-u') \;,
\end{equation}
where $\overline{\dots}$ means an average over realizations
of disorder with correlator $R_0(u)$. We will consider two types of disorder which correspond to two distinct 
universality classes: the so-called 
random-manifold (RM) class, and the random-periodic (RP) class.
In the RM class $R_0(u)$, the bare disorder correlator, is a rapidly decaying function over
distances larger than a short-distance cutoff $a$. 
In the RP class, the disorder has periodic correlations in the $u$ direction,
with a given periodicity $a_0$, 
such that $R_0(u+n a_0)=R_0(u)$ with $n$ an integer. 
For a given realization of the disorder the steady-state 
statistics of such interfaces is for $f=0$ described by a Boltzmann
weight $\propto \exp{[-\beta H]}$, with $\beta = T^{-1}$ the inverse temperature. 
For $f\neq 0$ and finite $T$ the system reaches a non-equilibrium 
moving steady-state at large enough times, while at $T=0$ it
can reach a moving steady-state only if the driving force is large enough to 
overcome the barriers between the disorder-induced 
metastable states.

Such disordered elastic systems (\ref{energy}) have been
widely studied during the last twenty years, in particular because they have
found many experimental realizations, ranging from
domain walls in ferromagnets \cite{domain_walls}, contact lines in wetting
\cite{moulinet_width, wetting} and fracture 
experiments \cite{fracture}. They are also relevant to describe periodic
structures like charge density waves \cite{cdw} or vortex lattices in type II
superconductors \cite{blatter, giamarchi}. In these systems, the competition
between elasticity and disorder leads not only to non-trivial ground state
configurations but also affects in a dramatic way their dynamical properties. In
particular, when driven by an external force $f$ at zero temperature, disorder
leads to a depinning transition at a
threshold value $f = f_c$, below which the interface is immobile, and above
which steady-state motion sets in. At finite but small $T$, 
an ultra-slow steady-state creep motion sets in below $f_c$, and 
in particular, at $f=0$, the zero velocity steady-state coincides 
with the equilibrium ground-state configuration when $T\to 0$.
Interestingly this model (\ref{energy})
has a universal roughness diagram~\cite{phasediagram} described 
by the crossovers between three types of self-affine interfaces, or ``reference steady-states'',
depending on the amplitude of the driving force $f$: 
\begin{itemize}
\item [(i)] at equilibrium ($f=0$) and $T=0$, which corresponds to the directed
polymer in a random medium, where the exponent $\zeta = 2/3$ for the RM class~\cite{kardar}, 
and $\zeta = 1/2$ in the RP class~\cite{rmtorp,solidonsolid},
\item [(ii)] exactly at the depinning threshold $f=f_c$ and $T=0$ where 
$\zeta \simeq 1.25$ in the RM class~\cite{rosso_1.25} 
and $\zeta = 3/2$ in the RP class~\cite{review_frg,rmtorp,rmtorp2}, 
\item [(iii)] in the limit $f \gg f_c$ where, in the moving frame, 
disorder induced fluctuations of the interface become 
an effective annealed noise: for the RM class this noise 
is thermal-like~\cite{chauve,phasediagram}, yielding the 
Edwards-Wilkinson roughness $\zeta=1/2$ corresponding to the 
thermal equilibrium of the system described by Eq.~(\ref{energy}) 
with $V(x,u) = 0$, while in the RP class it is a ``washboard'' 
colored noise yielding a larger exponent $\zeta = 1.5$, 
identical to the one of the RP depinning~\cite{rmtorp,rmtorp2}.
\end{itemize}

The third case (iii) effectively describes situations with annealed noise rather 
than quenched noise and in particular, for the RM class, the MRH distribution is simply 
given by the Airy distribution (\ref{scaling_airy}). In the two first cases (i) and (ii)
sample to sample fluctuations are important and one expects different fluctuations 
of the MRH than those produced by annealed noise. 
It is the purpose of this paper
to study the MRH distributions in the two first cases (i) and (ii) above, that 
we will denote by {\tt RMG}, {\tt RPG} for the random-bond (which we also call random-manifold) and random-periodic 
ground states, and {\tt RMD}, {\tt RPD} for the random-bond and random-periodic 
depinning configurations, respectively. These distributions are thus generated
by sample to sample fluctuations of the MRH, while the ensemble
of Gaussian signals ~(\ref{exp_Fourier}) can be physically thought as 
determined by the Boltzmann weight associated to~(\ref{eq:fractionalH}) with $T=1$.

We show that, in all the situations considered here, the distribution of the MRH
is very well described by the one corresponding to a Gaussian interface, where
$a_n$'s and $b_n$'s in Eq. (\ref{exp_Fourier}) are independent Gaussian
variables [as in Eq. (\ref{Gauss}) below] of zero mean and variances $\langle
a_n^2 \rangle = \langle b_n^2 \rangle \propto n^{-(1 + 2 \zeta)}$. Despite of
this, our numerical data show some numerical evidence that 
these distributions are different for the depinning configurations, and also 
for the ground-states in the random-manifold class.
Note that similar results were obtained for
the width distribution at the depinning threshold \cite{width_rosso}, which were
further justified using FRG calculations \cite{width_frg}. In that case, using
FRG close to the upper critical dimension, in dimension $d=4 -\epsilon$, it was
shown that the displacement field can be written as $u
= \sqrt{\epsilon} u_0 + \epsilon u_1$ where $u_0$ is a Gaussian random variable
of order ${\cal O}(1)$ while $u_1$ is a random variable with non-Gaussian
fluctuations, also of order ${\cal O}(1)$~\cite{width_frg}. Motivated by these similarities with
the Gaussian interface, we revisit the analysis of the right tail of the pdf of
the MRH for Gaussian interfaces and generic $\zeta$ where we obtain exact
results,  using Pickands'~theorem~\cite{pickands}. 
For the ground-states in the random-periodic class with periodic boundary conditions 
we find that corrections to the Gaussian independent modes approximation vanish at large system 
sizes and thus the MRH distribution becomes indistinguishible from the 
universal Airy distribution. Finally, for the equilibrium case we show
that the MRH distribution is sensitive to the boundary conditions, 
highlighting their importance in the study of anomalous, self-affine interfaces.

The paper is organized as follows: in Section II, we remind the results that
were obtained for Gaussian interfaces, and derive the precise asymptotic
behavior of the right tail of the distribution of the MRH. In Section III, we
provide the details of our numerical simulations, the results of which are then
presented in Section IV. In Section V, we discuss these results before we
conclude in Section VI. Some details relative to the use of Pickand's theorem
have been left in appendix A.

\section{Results for Gaussian interfaces}

We first start to remind the known results for Gaussian interfaces, without disorder. The distribution of $h_{m}$ was first studied for the simplest
model of an elastic interface described by Eq. (\ref{energy}) with $V = 0$, $f=0$ and periodic boundary conditions, {\it i.e.} $u_L = u_0$. In that case one has ${\rm w}_L^2 = T L/12$. While a first numerical study indicated that $h_m \sim L^{\zeta}$, with $\zeta = 1/2$, Majumdar and Comtet were then able to compute exactly the full distribution of $h_m$~\cite{satya_airyshort, satya_airylong}, $P(h_m|L)= \frac{1}{L^{1/2}}\, f_{\rm Airy}\left(\frac{h_m}{L^{1/2}}\right)$
where $f_{\rm Airy}$ is the so-called {\it Airy distribution}, whose Laplace transform is given by
\begin{eqnarray}
&& \int_0^{\infty} f_{\rm Airy}(x) e^{-sx} dx = s\sqrt{2\pi} \sum_{k=1}^{\infty}
e^{-\alpha_k 
s^{2/3}2^{-1/3}} ,
\label{lt1}
\end{eqnarray} 
where $\alpha_k$'s are the amplitudes of the zeros of the Airy function on the negative real axis. For instance, $\alpha_1=2.3381\dots$, $\alpha_2=4.0879\dots$, $\alpha_3=5.5205\dots$ etc~\cite{abramowitz}. It is also possible to invert this Laplace transform to obtain~\cite{satya_airyshort, satya_airylong, takacs_invert}
 \begin{equation}
f_{\rm Airy}(x) = \frac{2\sqrt{6}}{x^{10/3}}\sum_{n=1}^\infty e^{-b_n/x^2}
b_n^{2/3} U(-5/6,4/3,b_n/x^2) \;,
\label{Airy_dist}
\end{equation}
where $b_n = 2\alpha_n^3/27$ and $U(a,b,z)$ is the confluent hypergeometric (Tricomi's) function~\cite{abramowitz}. Its asymptotic behavior
is given, for small argument by
\begin{eqnarray} \label{airy_smallx}
f_{\rm Airy}(x) &\sim & \frac{8}{81} \alpha_1^{9/2} x^{-5}\, e^{-
2\alpha_1^3/{27 x^2}} \quad {\rm
  for}\quad x\to 0  \;.
\end{eqnarray}
The large argument behavior was obtained in Refs.~\cite{satya_airylong, janson_louchard}
\begin{eqnarray}\label{airy_largex}
f_{\rm Airy}(x) \sim {72}{\sqrt{\frac{6}{\pi}}} x^2 e^{-6 x^2} \quad {\rm for}\quad x\to \infty \;.
\end{eqnarray}
In the above expression, the leading Gaussian behavior, $e^{-6x^2}$, was obtained in Ref. \cite{satya_airylong} while the subleading prefactor, ${72}{\sqrt{\frac{6}{\pi}}} x^2 $ was obtained in Ref. \cite{janson_louchard} using rather involved combinatorial techniques [which also allow to obtain the subleading corrections to (\ref{airy_largex})]. We show below that 
this result (\ref{airy_largex}) can be straightforwardly obtained using Pickands' theorem~\cite{pickands} concerning the asymptotic properties of the maximum of a stationary Gaussian (see also appendix \ref{appendix_A} for more details). 

The MRH distribution was then investigated for Gaussian interfaces displaying a $1/f^\alpha$ power spectrum in Ref.~\cite{GMOR}. Considering again periodic boundary conditions, $h(0) = h(L)$, the relative height field can be written in a Fourier expansion as in Eq. (\ref{exp_Fourier}) with a 
Gaussian probability measure: 
\begin{eqnarray}\label{Gauss}
{\cal P}[h] \propto \exp \left[-\frac{1}{4}\sum_{n=1}^\infty \frac{(2\pi n)^\alpha}{L^{\alpha-1}} (a_n^2 + b_n^2) \right] \;,
\end{eqnarray}
corresponding to independent Gaussian random variables $a_n$'s and $b_n$'s of zero mean and variance, for $n \geq 1$ 
\begin{eqnarray}\label{variance}
\langle a_n a_{n'} \rangle = \langle b_n b_{n'} \rangle = \delta_{n,n'}\frac{2L^{\alpha-1}}{(2 \pi n)^\alpha} \;, 
\end{eqnarray}
while $\langle a_n b_{n'} \rangle = 0$. Note that this is the same convention as the one chosen in Ref. \cite{GMOR}, which yields in particular
\begin{equation}
 {\rm w}^2_L = \frac{2L^{\alpha-1}}{(2 \pi)^{\alpha}} {\bm \zeta}(\alpha) \;, \; {\bm \zeta}(\alpha) = \sum_{n=1}^\infty n^{-\alpha} \;,
\end{equation}
where ${\bm \zeta}(\alpha)$ is the Riemann zeta function. The roughness exponent is thus $\zeta = (\alpha-1)/2$. In Ref. \cite{GMOR} it was shown that the distribution of the MRH takes the scaling form
\begin{eqnarray}\label{gauss_scaling}
 P(h_m|L) = \frac{1}{L^{\zeta}} \tilde f_{\alpha}\left( \frac{h_m}{L^\zeta}\right) \;,
\end{eqnarray}
where from now on, the tilde over any 
distribution function indicates that it corresponds to Gaussian interfaces described by Eq. (\ref{Gauss}). While the exact expression of the scaling function $\tilde f_\alpha(x)$
is known only for $\alpha = 2$, see Eqs. (\ref{scaling_airy}, \ref{Airy_dist}). Nevertheless, for $\alpha = 2n$ where $n \in \mathbb{N}$, the computation of the $P(h_m|L)$ can be formulated in terms of a path integral \cite{GMOR} from which several properties can be deduced. In particular, the small argument behavior of the scaling function can be studied in detail, yielding~\cite{GMOR}
\begin{eqnarray}\label{gauss_small}
 \tilde f_{\alpha}(x) \sim C x^{-\gamma} \exp{(-B/x^\beta)} \;,
\end{eqnarray}
with 
\begin{eqnarray}\label{beta_gamma}
 \beta = \frac{2}{\alpha -1}=\frac{1}{\zeta} \;, \; \gamma = \frac{2 \alpha + 1}{\alpha - 1} = \frac{3 + 4\zeta}{2 \zeta} \;,
\end{eqnarray}
while the amplitudes $B$ and $C$ can be expressed in terms of the smallest eigenvalue $\epsilon_0 \equiv \epsilon_0(\alpha) > 0$ of a linear eigenvalue problem \cite{GMOR}:
\begin{equation}\label{gauss_ampli_small}
 B = \frac{\alpha-1}{2} \left(\frac{2 \epsilon_0}{\alpha + 1} \right)^{\frac{\alpha+1}{\alpha-1}} \;, \; \sqrt{\frac{\alpha+1}{\alpha-1}} \left(\frac{2 \epsilon_0}{\alpha+1} \right)^{\frac{3}{2}\frac{\alpha+1}{\alpha-1}} \;.
\end{equation}
Note that $\epsilon_0(2) = \alpha_1/2^{1/3}$ so that these formulas (\ref{gauss_small}, \ref{beta_gamma}, \ref{gauss_ampli_small}) coincide with the small argument behavior of the Airy function for $\alpha = 2$ (\ref{airy_smallx}). While this behavior (\ref{gauss_small}) where the exponents are given by Eq. (\ref{beta_gamma}), was shown to hold for $\alpha = 2n$, numerical simulations demonstrated its validity for any $\alpha > 1$ \cite{GMOR}. We also notice that the value of this eigenvalue $\epsilon_0(\alpha = 4)$, corresponding to the Random Acceleration Process, was shown to appear in a completely different problem of random convex geometry, called the Sylvester's question \cite{hilhorst_sylvester}. 

The large argument behavior of the scaling function $\tilde f_\alpha(x)$ (\ref{gauss_scaling}) is harder to treat analytically. Numerical simulations performed in Ref. \cite{GMOR} showed evidence for the asymptotic behavior:
\begin{eqnarray}\label{gauss_large}
 \tilde f_{\alpha}(x) \sim D x^{\delta} e^{-E x^2} \;.
\end{eqnarray}
As explained in the Appendix \ref{appendix_A}, we are able to compute this exponent $\delta$ as well as the amplitudes $D$ and $E$ using Pickands'~theorem. One obtains that the exponent $\delta$ depends on $\alpha$ as follows
\begin{eqnarray}\label{delta}
 \delta = 
\begin{cases}
&\frac{2}{\alpha-1} \;, 1 < \alpha < 3 \\
&1 \;, \alpha > 3 \;,
\end{cases}
\end{eqnarray}
while one expects logarithmic corrections for $\alpha =3$,which thus appears as a threshold value. These values are consistent with the numerical
estimations reported in Ref. \cite{GMOR}. They are also consistent with the exact result obtained in the limit $\alpha \to \infty$, for which $\delta = 1$ \cite{GMOR}. The amplitude $E$ is given by
\begin{eqnarray}\label{ampli_E}
 E = \frac{(2 \pi)^\alpha}{4 {\bm \zeta}(\alpha)} \;,
\end{eqnarray}
while the amplitude $D$ can be expressed in terms of a Pickands' constant (\ref{pickands_constant}), and has in general a complicated expression. However for $\alpha = 2$ and $\alpha > 3$ it can be computed explicitly as
\begin{eqnarray}\label{ampli_D}
D = 
\begin{cases} 
& 72 \sqrt{\frac{6}{\pi}} \;, \; {\rm if} \; \alpha = 2 \;, \\
&\sqrt{\frac{{\bm \zeta}(\alpha-2)}{{\bm \zeta}(\alpha)}} \frac{(2 \pi)^\alpha}{2 {\bm \zeta}(\alpha)}  \;, \alpha > 3 \;.
\end{cases}
\end{eqnarray}
The expression of $D$ for generic $1 < \alpha < 3$ is left in Eq. (\ref{gen_alpha}) in Appendix A. It is straightforward to check that the results in Eqs. (\ref{gauss_large}, \ref{delta}, \ref{ampli_E}, \ref{ampli_D}) yield back, for $\alpha=2$, the result for the Airy distribution (\ref{airy_largex}).

\section{Various situations for elastic interfaces in disordered media : details of simulations}

We now describe the various models of elastic interfaces that we study using numerical simulations. Numerically, it is very convenient to discretize the internal $x$ direction of the elastic strings described by Eq.~(\ref{energy}). 
Indeed, once conveniently discretized, the ground-states and critical states at depinning can be obtained 
by exact and fast numerical algorithms. As discussed below, for depinning we can keep $u(x)$ as a continuous variable. 
For the ground-state it is convenient to also discretize $u(x)$ and work on a lattice. 
None of these practical choices change the universality classes we analyze, and at large 
enough $L$ the numerical results for the lattice model should be indistinguishable from those of 
a continuum model.

\subsection{The ground state}
When the external force is zero and the finite size pinned interface is allowed to equilibrate 
at zero temperature it reaches the ground state configuration, found 
by minimizing the energy~(\ref{energy}) for a given sample of disorder.
Finding such a state for each disorder realization 
poses a global optimization problem which can be solved by an exact transfer-matrix 
method. This method allows to find the ground-state of a line on a lattice where both $u(x)$ and $x$ are 
discrete variables. For simplicity we impose a hard constraint on the local allowed elongations, 
$|u(x+1)-u(x)|\leq 1$. This simplification does not change the equilibrium universality class which is 
thus only determined by the disorder correlator. This is at variance with depinning configurations, 
where such constraints, or more general non-linear corrections to the standard harmonic elasticity, 
can change the depinning universality classes~\cite{rosso_algorithm,phasediagram}. We thus simulate the directed polymer in a random 
medium as a simple model for the continuum interfaces described by~(\ref{energy}), with the 
universality classes determined by the nature of the discretized disorder potential correlations. 

We work on a finite discrete space: we use a rectangular lattice of size $L\times \ell$.
Points of the lattice are indexed by the couple of integers $(i,j)$ in the ranges $0 \leq i \leq L$ and $0\leq j \leq \ell$.
The disordered potential is now a random variable $V_{ij}$ given at each point $(i,j)$ of the described lattice, 
with zero mean
\begin{subequations}
\begin{equation}
\overline{V_{ij}}=0,
\end{equation}
and correlations
\begin{equation}
\overline{V_{ij} V_{kl}}= \delta_{ik} \delta_{jl} \;.
\end{equation}
\end{subequations} 
In our simulations, we choose Gaussian random variables,
although the precise distribution should not be relevant, 
provided that it remains narrow.
We work with periodic boundary conditions {\it both} in the $u$ and in the $x$ direction: 
the sites $(0,j)$ are identified with the sites $(L,j)$, for all $0 \leq j \leq \ell$,
and in the $u$ direction as well: $(i,0)\equiv (i,\ell)$, for all $0\leq i \leq L$.
A path $\mathcal{C}$ of length $L$ is given by the sequence of visited sites $(i,j)$,
for $0 \leq i \leq L-1$, and due to periodic boundary conditions, it can be regarded as 
a close path on a torus. 
Its energy is the sum of disordered energies at each visited site:
\begin{equation}
E(\mathcal{C})= \sum_{(i,j) \in \mathcal{C}} V_{ij} \;.
\end{equation}
At each step, the path can go from $(i,j)$ to one of the three possibilities
$(i+1,j-1)$, $(i+1,j)$ or $(i+1,j+1)$:
we then select the optimal path amongst the ensemble $\mathcal{C}_{j_0}$ of paths
starting in $(0,j_0)$ and ending in $(L,j_0) \equiv (0,j_0)$ by a recursive method~\cite{kardar}. Calling $\mathcal{C}^*_{j_0}$ this path, one has
\begin{equation}
E_{\text{opt}}(j_0)= E(\mathcal{C}^*_{j_0}) = \min_{\mathcal{C}_{j_0} } E( \mathcal{C}_{j_0}) \;, 
\end{equation}
the energy for the optimum path, minimizing the total energy for a fixed starting point $j_0$.
However considering only these paths $\mathcal{C}^*_{j_0}$, even averaging over the disorder, yields
a non-stationary two-point correlation function of the displacement field. One obtains 
a stationary correlation function by considering the optimal path $\mathcal{C}^*$ among all closed paths 
on the torus. Then
\begin{equation}
E_{GS}=\min_{\mathcal{C}} E(\mathcal{C})=E(\mathcal{C}^*) .
\end{equation}
From our ensemble of paths $\mathcal{C}^*_{j_0}$, one has to minimize by varying the starting point $j_0$
on the $u$ axis:
\begin{equation}
E_{\textrm{GS}}= \min_{0\leq j_0 \leq \ell} E_{\textrm{opt}}(j_0) ,
\end{equation}
Recalling that $L$ and $\ell$ are the lengths of the disordered substrate in the $x$ and $u$ directions 
respectively, one has to choose $\ell  \gtrsim L^{\zeta_{\textrm{GS}}}$ in order to explore a region of correlated paths.
In this case, $\zeta_{\textrm{GS}}$ stands for the ground state roughness exponent, which can take 
the values $\zeta_{\tt RMG}=2/3$ \cite{kardar} for the random-mainfold ground state and $\zeta_{\tt RPG}=1/2$ 
for the random-periodic ground state, respectively.
In our simulations, we have taken $\ell=L$ to produce ground states in the RM universality class, 
while we took $\ell \ll L$, for ground states in the RP universality class. The last choice is equivalent to construct 
a quenched potential with periodic correlations in the direction of $u$ with period $\ell$.
In particular, if we take $\ell \ll L$ but still large compared to the unity, $\ell$ controls the crossover between the 
RM and RP universality classes, from $\zeta_{\tt RMG}$ to $\zeta_{\tt RPG}$. For this reason, in most 
of our simulations for the RP class, we set $l\sim O(1)$, 
thus minimizing the crossover or finite-size effect.

We have produced a number of $10^6$ samples of each lengths $L=128, 256, 512, 1024, 2048, 4096$
in order to construct the histograms of the maximal relative height $h_m$.
The results we obtain for the MRH and discuss in the next section are computed from 
stationary paths $\mathcal{C}^*$, and are referred with the index {\tt RMG}, {\tt RPG} for the random manifold and random 
periodic universality classes. 
We discuss the results for the ensemble given by $\mathcal{C}^*_{j_0}$ (with the arbitrary choice $j_0=0$)
having a non-stationary correlation function in the section~\ref{boundary}, for the random-manifold case 
using the index {\tt RMG'} to distinguish from its stationary counterpart {\tt RMG}. 
Both cases, {\tt RMG} and {\tt RMG'}, have the same roughness exponent $\zeta_{\tt RMG}=\zeta_{\tt RMG'}=2/3$.

\subsection{The depinning transition}
The critical configuration $u_c(x)$ of an elastic interface 
at the depinning transition is an extreme solution of the overdamped 
equation of motion
\begin{equation}
\label{eq:metasol}
\frac{\partial u(x,t)}{\partial t} = 
-\frac{\delta H[\{u(x)\}]}{\delta u(x)} + f = 0,
\end{equation}
such that $u = u_c$ for  
the largest possible force $f=f_c$ satisfying the above equation, with $f_c$ the 
so-called critical force. 
Above $f_c$, the sum of all forces in the second term of Eq.~(\ref{eq:metasol}) 
is always positive, and thus a zero velocity steady-state solution can not exist. 
Middleton's theorems~\cite{rosso_1.25} assure that such a solution exists 
and it is unique. 
In order to solve numerically Eq.~(\ref{eq:metasol}) for a given realization of 
the random potential we can discretize the string in the direction $x$ 
into $L$ pieces, thus converting Eq.~(\ref{eq:metasol}) into an inversion problem.
Solving the resulting $L$-dimensional system of Eq.~(\ref{eq:metasol}) for large $L$ 
by standard general methods is however a formidable task, due to the  
non-linearity of the pinning force. On the other hand, solving the long-time steady-state dynamics at 
different driving forces $f$ both below and above $f_c$ is very inefficient due to the 
critical slowing down near $f_c$. Fortunately, this problem has a particular analytical structure 
that allows to devise a precise and very efficient algorithm allowing to obtain 
iteratively the critical force $f_c$ 
and the critical configuration $u_c(x)$ for each independent disorder realization without 
solving the actual dynamics nor directly inverting the discretized version of 
Eq.~(\ref{eq:metasol}) ~\cite{rosso_algorithm,rosso_1.25}. In this paper we use such method 
with periodic boundary conditions in all directions. This guarantees that the 
critical configurations have spatially stationary correlation functions.

We implement the algorithm to find the critical configuration 
in 1+1 dimensions
as in Ref.~\cite{rosso_1.25}.
We discretize the space in the $x$ direction while keeping 
$u$ as a continuous variable. A potential $V(x,u)$ 
satisfying (\ref{eq:potential}) is modeled with random cubic splines. 
We consider periodic boundary conditions in both directions, in a system 
of size $L \times \ell$. When $\ell$ is large enough, the critical 
configuration sample averaged width is well described by 
${\rm w}_L \equiv [\overline{{u_c}^2-\overline{u_c}^2}]^{1/2} 
\sim L^{\zeta}$ 
with $\zeta \equiv \zeta_{\tt RMD} \approx 1.25$~\cite{rosso_1.25} the random-manifold 
depinning roughness exponent. We will denote this Random-Manifold depinning case as {\tt RMD}. 
For small $\ell$ the average width is well described by the random-periodic depinning
roughness exponent 
$\zeta \equiv \zeta_{\tt RPD} \approx 1.5$~\cite{review_frg,rmtorp,rmtorp2}. 
Accordingly, we will denote this Random-Manifold depinning case as {\tt RPD}. 
More precisely, for this system it was found that 
${\rm w}_L \sim G(\ell/L^{\zeta_{\tt RMD}}) L^{\zeta_{\tt RMD}}$ 
for all values of $\ell$ with $G(x)$  
a non-monotonic function of $x$,  such that 
$G(x)\sim x^{(1-\zeta_{\tt RPD}/\zeta_{\tt RMD})}$ for $x\ll 1$ and 
a with a very slow, approximately logarithmic, growth for $x\gg 1$~\cite{rmtorp2}.
Changing the transverse periodicity thus allows to crossover from the random-manifold 
to the random-periodic universality class. By using this method we can sample 
critical configurations belonging to these two classes, 
{\tt RMD} and {\tt RPD}, respectively, and tune the non-universal prefactors 
${\rm w}_L/L^{\zeta_{\tt RMD}}$, ${\rm w}_L/L^{\zeta_{\tt RPD}}$ to different values. Once the 
critical configuration $u_c(x)$ 
is obtained for each case, we substract the center of mass position 
$h(x)=u_c(x)-(1/L)\int_0^L dx'\;u_c(x')$ and calculate the MRH from 
Eq. (\ref{eq:MRH}). Repeating this procedure for different disorder 
realizations gives access to the MRH probability distribution. 
In our simulations we use between $10^4$ and $10^5$ critical samples of sizes $L=128,256,512$.

\section{Numerical results}

We present numerical results of the probability distribution function $P_k(h_m|L)$ of the maximal relative height $h_m$ 
where the subindex $k={\tt RMG, RPG, RMD, RPD}$ denotes the different cases: random-manifold and 
random-periodic ground states, and random-manifold and random-periodic critical 
configurations, respectively.

\subsection{Scaling of $h_m$ with $L$}

\begin{figure}[h]
\begin{center}
\includegraphics[width=\linewidth]{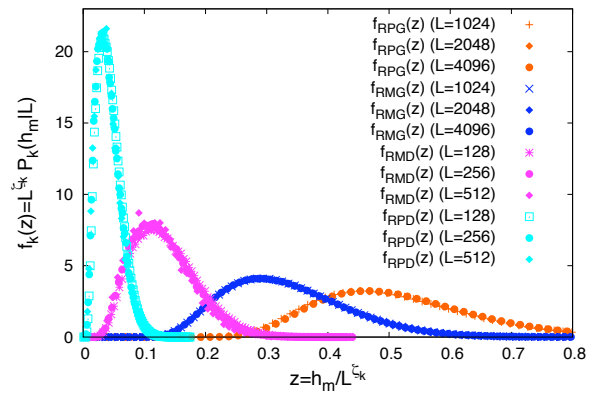}
\caption{For the random-manifold and random-periodic ground states, $k={\tt RMG, RPG}$, and 
critical configurations $k={\tt RMD, RPD}$, we plot the rescaled distributions $f_k(z)=L^{\zeta_k} P_k(z\times L^{\zeta_k}|L)$,
obtained from the histograms $P_k(h_m|L)$ computed for different system size $L$ as indicated in the key.}
\label{fig_length-scaling}
\end{center}
\end{figure}

For all cases, our numerical data are compatible with the scaling law
\begin{equation}
\label{length-scaling}
P_k(h_m|L)=\frac{1}{L^{\zeta_k}} \ f_k\left( \frac{h_m}{L^{\zeta_k}} \right),
\end{equation}
where $\zeta_k$ is the roughness exponent of the case $k$, $\zeta_{\tt RMG}=2/3$, $\zeta_{\tt RPG}=1/2$, 
$\zeta_{\tt RMD}\approx 1.25$, and $\zeta_{\tt RPD} = 3/2$.
In the continuum limit $L\gg 1$, 
the rescaled functions $f_k$  are expected to depend only on the rescaled maximal relative height $z=h_m/L^{\zeta_k}$.
In Fig.~\ref{fig_length-scaling} one observes that the rescaled distributions for different $L$ of the case $k$ indeed 
collapse for different sizes of the same class, strongly supporting this scaling relation~(\ref{length-scaling}).
The different cases, {\tt RPG, RMG, RMD} and {\tt RPD} in order of roughness exponent, 
have very different scaling functions $f_k$ in this $L^\zeta$ scaling.
In particular, it is clearly visible that the most probable value, the mean and also 
the standard deviation of each $f_k$ decreases when the roughness exponent increases.
Besides these facts, the curves are somewhat similar, and we are interested in 
testing other simple scaling forms to compare these distributions $P_k(h_m|L)$ on the same footing.

\subsection{Average and sigma scaling}

\begin{figure}[h]
\begin{center}
\includegraphics[width=\linewidth]{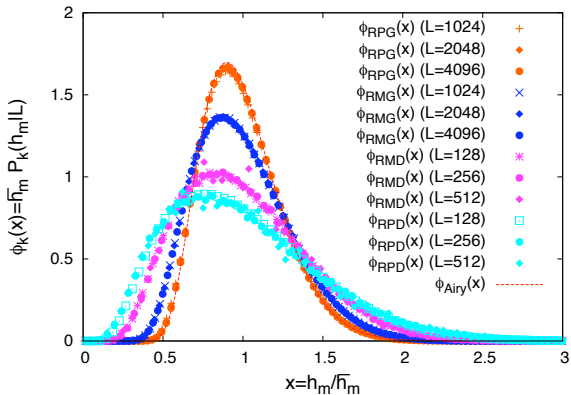}
\caption{For each case $k={\tt RPG, RMG, RMD, RPD}$ we show the average-rescaled 
distributions $\phi_k(x)=\overline{h_m} P_k(x \overline{h_m} |L)$.
Different system lengths are shown, as quoted in the key,
and the rescaled functions collapse onto the same curves, having unit average.}
\label{fig_average-scaling}
\end{center}
\end{figure}

\begin{figure}[h]
\begin{center}
\includegraphics[width=\linewidth]{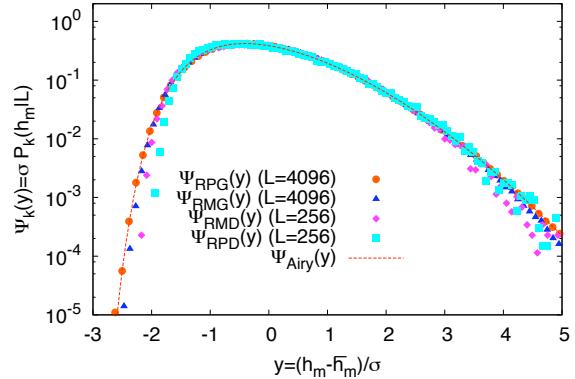}
\caption{For each case $k={\tt RPG, RMG, RMD, RPD}$ we show the sigma rescaled distributions
 $\psi_k(y)= \sigma P_{k}(\overline{h_m} + \sigma y |L)$,
obtained with $L=256$ for the depinning cases and $L=4096$ for the ground state cases. Other curves for different $L$ collapse on theses ones and are not shown for clarity.
The Airy distribution is shown also with sigma scaling.
}
\label{fig_sigma-scaling}
\end{center}
\end{figure}

Two other scaling forms have been used in the literature~\cite{GMOR,width_rosso}, namely the {\it average-scaling} and 
the {\it sigma-scaling}. The average scaling is defined by
\begin{equation}
P_k(h_m|L)=\frac{1}{\overline{h_m}} \ \phi_k\left(\frac{h_m}{\overline{h_m}}\right) ,
\end{equation}
where the average is computed as
\begin{equation}
\overline{h_m}=\int_0^\infty \rmd h_m \, h_m \ P_k(h_m|L) .
\end{equation}
The rescaled variable $x=h_m/\overline{h_m}$ has by definition a unit mean.
It differs from the precedent scaling by the non-universal prefactor $A_k$ in $\overline{h_m} =A_k L^{\zeta_k}$. 
This prefactor can be indeed physically relevant if there exists a crossover to 
the asymptotic roughness regime which characterizes the universality class. In such cases, 
if the crossover to the asymptotic regime takes place at a characteristic length $L_{\times}$, we 
can expect $\overline{h_m} \approx A_{\times} (L/L_{\times})^{\zeta_k}$, 
with $A_{\times} \sim \overline{h_m(L_{\times})}$ a characteristic maximal height for a system 
of size $L_{\times}$. 
We thus obtain $A_k \approx A_{\times}/L_{\times}^{\zeta_k}$. Such crossovers usually can depend 
on microscopic details, such as the strength of the disorder and the elasticity 
({\it e.g.} the Larkin length), temperature or the spatial discretization when the interface 
is defined on a lattice. In the case of moving interfaces such crossovers can 
in addition depend on the velocity~\cite{phasediagram}. As a peculiar case, 
in very elongated samples the prefactor $A_k$ could also depend on the 
transverse dimension of the system if the configuration carries with it an 
extreme value over the sample, such as the depinning threshold for the 
critical configuration~\cite{rmtorp2}. We present this average scaling in Fig.~\ref{fig_average-scaling}. 
We see again a very good collapse for the different system sizes within each case $k$, 
and also observe that the scaling functions $\phi_k$ look still very different 
because the width of the rescaled distributions $\phi_k$ appreciably 
increases when the roughness exponent increases.
\par
Yet another way to present the data is the sigma-scaling defined by
\begin{equation}
P_k(h_m|L)=\frac{1}{\sigma} \, \psi_k \left( \frac{h_m-\overline{h_m}}{\sigma} \right),
\end{equation}
where the standard deviation $\sigma$ is obtained when averaging over the disorder
\begin{equation}
\sigma^2=\overline{\left(h_m-\overline{h_m}\right)^2} .
\end{equation}
The rescaled variable $y=(h_m-\overline{h_m})/\sigma$ has zero average and unit standard deviation.
This is the most general linear transformation that can get rid of model dependent amplitudes, such as 
the aforementioned $A_k$.
It allows then to compare the shape of the distributions on the same footing.
Actually, the differences between the distributions $\psi_k$ are minute, and not visible on a plot with linear axis.
In Fig.~\ref{fig_sigma-scaling} we show the sigma-rescaled distributions $\psi_k$ obtained from samples with $L=4096$ and $L=256$ for ground 
states and depinning respectively, using log-linear axis (distributions obtained from other system sizes collapse on 
the same curves and are not shown for clarity). 
One can observe that the $\psi_k$'s are eventually different: this is especially visible in the left tail.
It indicates that the rescaled distributions remain sensitive to the roughness exponent $\zeta$. The Airy distribution function (also plotted in sigma scaling) 
seems to collapse very well with the $\texttt{RPG}$ case (see the discussion in paragraph~\ref{comparison-gaussian} below).
Moreover it gives an approximate description of the remaining cases \texttt{RMG}, \texttt{RMD}, \texttt{RPD},
furnishing an idea of the small and large argument behaviors. In particular we have checked that the large argument behavior of $\psi_k(y)$, for each case
$k={\tt RPG, RMG, RMD, RPD}$, is well fitted by a Gaussian tail, $\psi_k(y) \propto \exp(-d_k y^2)$, although a precise determination of the coefficients $d_k$ remains a hard task numerically. Considering this observation together with our exact analytical results for Gaussian interfaces obtained above (\ref{gauss_large}), we
conjecture that $\psi_k(y) \propto \exp(-d_k y^2)$ is the exact leading behavior of $\psi_k(y)$ for large $y$.

\subsection{Comparison to Gaussian Independent Modes}
\label{comparison-gaussian}

\begin{figure}[h]
\includegraphics[width=\linewidth]{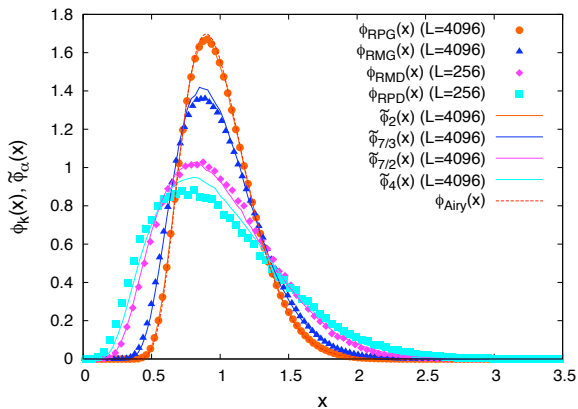}
\caption{Comparison between the distributions $\phi_k(x)$ (symbols) of the MRH of the elastic interfaces in random media,
for each case $\zeta_k=1/2, 2/3,1.25,1.5$ for $k={\tt RPG, RMG, RMD, RPD}$, obtained from the length $L=4096,4096,256,256$, 
and the distributions {${\tilde \phi}_{\alpha}(x)$ (lines) of the MRH of Gaussian interfaces
with $\alpha=2,7/3,7/2,4$} and corresponding roughness exponents $\zeta_k$. All distributions are shown with average scaling.
The Airy distribution, corresponding to $\zeta=1/2$, is plotted as a reference. }
\label{fig_gauss}
\end{figure}

\begin{figure}[h]
\includegraphics[width=\linewidth]{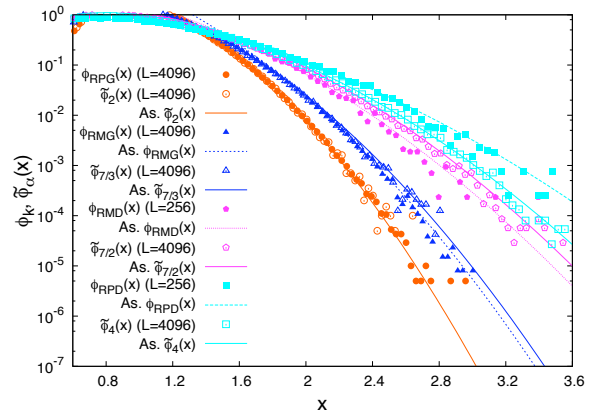}
\caption{{Right tails of the MRH distributions in average scaling. We show the data for Gaussian interfaces (open signs)
with corresponding values $\alpha=2,7/3,7/2,4$, 
and their exact asymptotics (continuous lines) computed from Eq.~(\ref{gauss_large}).
We also show the disordered interfaces (filled signs) with $\zeta_k$ for the
four studied cases with $k={\tt RPG, RMG, RMD, RPD}$. For each case 
the fitted asymptotic tails as $\exp (-e_k x^2+c_k)$ are also presented (discontinuous lines).}}
\label{fig_tails}
\end{figure}

A natural question arises when studying disordered systems: 
is there a simpler model that can absorb the complexity induced by the spatial randomness? If a satisfying answer comes out, it will permit to obtain more information on the disordered system.
Indeed, models without disorder are often easier to work on numerically, 
and are also more tractable analytically. 
For our problem a good candidate is the Gaussian independent mode interface as it can easily describe 
the self-affine geometry of the pinned interface. Physically, the sample 
to sample MRH fluctuations due to different disorder realizations are thus 
approximated by the ``thermal'' MRH fluctuations of a free elastic 
interface with a non-local elasticity as described by Eq.~(\ref{eq:fractionalH}). 
Comparing with Gaussian signals hence allows to separate the purely geometrical features 
of self-affine disordered interfaces, 
which can be successfully described by independent Fourier modes, from the specific non-Gaussian 
corrections generally expected from the complex interplay 
between disorder and elasticity. This analysis is thus experimentally relevant as 
it can provide very specific information from a purely geometrical analysis of 
interfaces.

Hence we compare our results to the distribution of MRH obtained with Gaussian interfaces 
generated via the Fourier expansion~(\ref{exp_Fourier}) with the probability measure given in~(\ref{Gauss}). 
We therefore adjust the parameter $\alpha$ which parameterizes the Gaussian signal
to get the corresponding roughness exponent. Since the roughness exponent of the Gaussian interface is
$\zeta = (\alpha-1)/2$ we will use the notation $\alpha_{\tt RPG}=2$, $\alpha_{\tt RMG}=7/3$, $\alpha_{\tt RMD}=7/2$, and $\alpha_{\tt RPD}=4$ to characterize the Gaussian cases used to compare with the corresponding disordered interfaces.
Average and sigma scaling of the MRH distribution 
is used to adjust the amplitude of the Gaussian interface (or ``temperature'') and 
get the best Gaussian approximation for the ensemble of pinned interfaces.

To start the comparison, we have computed the MRH distribution ${\tilde \phi}_{k}$
in average scaling, for the 
Gaussian signals corresponding to the four cases. In Fig.~\ref{fig_gauss} 
one can see that the distributions for the elastic interfaces in disordered media are 
well approximated by their pure Gaussian counterpart, especially for $k={\tt RPG},{\tt RMG}$. 
In Fig.~\ref{fig_tails}, we zoom on the large $x$ behavior in a log-linear axis, and plot the exact asymptotics
for Gaussian interfaces which we computed above in Eqs.~(\ref{gauss_large}, \ref{delta}, \ref{ampli_E}, \ref{ampli_D}) -- up to a rescaling of $h_m$ -- and the fitted asymptotics for the disordered interfaces.
As mentioned in the analysis in sigma scaling of Fig.~\ref{fig_sigma-scaling}, the data shows a 
good agreement with a Gaussian tail $\exp(-e_k x^2)$, at least at leading order. One sees 
that the coefficients $e_k$ are slightly different from those entering the exact asymptotics of 
Gaussian interfaces. We also observe that the value of $e_k$ is closer to its Gaussian 
counterpart for the case of ground-states, {\tt RPG}, {\tt RMG}.

\begin{figure}[h]
\includegraphics[width=\linewidth]{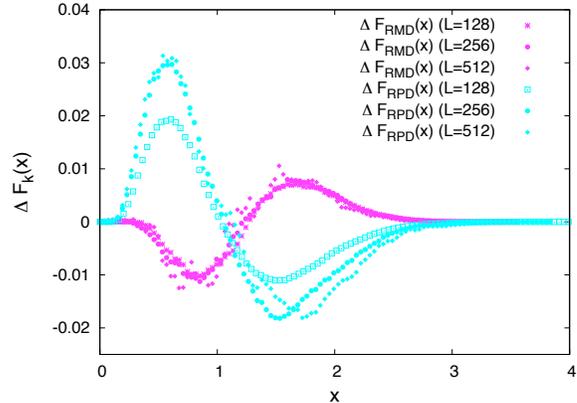}
\caption{Differences $\Delta F_k(x)$ of cumulative distributions between disordered interfaces and Gaussian 
interfaces, for the depinning cases $k={\tt RMD, RPD}$, for system lengths {$L=128,256,512$ for the disordered interfaces data}, 
in average scaling.}
\label{fig_diff-cumul-dep}
\end{figure}

\begin{figure}[h]
\includegraphics[width=\linewidth]{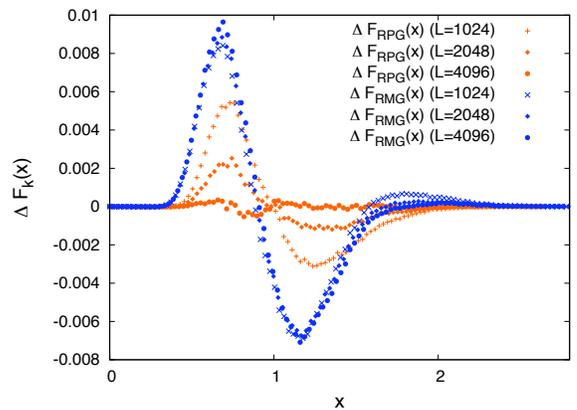}
\caption{Differences $\Delta F_k(x)$ of cumulative distributions between disordered interfaces and Gaussian 
interfaces, for the ground-state cases $k={\tt RPG, RMG}$, for system lengths {$L=1024, 2048, 4096$
for the disordered interface data}, 
in average scaling.}
\label{fig_diff-cumul-gs}
\end{figure}

To go further and characterize better the possible differences with Gaussian signals we proceed  
as in Ref.~\cite{width_rosso} for the width distribution and compute (numerically) the cumulative 
MRH distributions for each $k={\tt RPG, RMG, RMD, RPD}$ {for different sizes $L$}
and their corresponding 
Gaussian counterparts (denoted with the tilde) 
{as references, computed with the largest size $L=4096$ to minimize finite-size effects:
}
\begin{align}
F_k(x)&=\int_0^x \rmd x' \ \phi_k(x'), \\
{\tilde F}_{k}(x)&=\int_0^x \rmd x' \ {\tilde \phi}_{\alpha_k}(x'), \ L=4096,
\end{align}
and analyze the difference $\Delta F_k(x)=F_{k}(x)-{\tilde F}_{k}(x)$ for different values of the length $L$.
In Fig.~\ref{fig_diff-cumul-dep} we can see that for the depinning cases, {\tt RMD} and {\tt RPD}, 
the differences $\Delta F_k(x)$ seem to saturate for all values 
of the rescaled variable $x=h_m/\overline{h_m}$. This is consistent with 
what was found for the same difference regarding the width 
distribution for critical configurations at depinning~\cite{width_rosso}.
In Fig.~\ref{fig_diff-cumul-gs} we show the ground-states cases.
While the {\tt RMG} case maintains a finite difference as $L$ increases
for the {\tt RPG} the differences {decrease towards zero} for all $x$ 
with increasing size. 

To quantify the difference between the MRH distributions of elastic interfaces 
in random media and their corresponding Gaussian signal we start with  
a very simple statistical test. We assume that disordered elastic 
interfaces can be mapped onto a Gaussian signal 
by a simple rescaling of its amplitude 
by a factor $\chi$, 
\begin{equation}
 h(x) = \chi {\tilde h(x)},
\label{eq:hypothesis}
\end{equation}
and thus the two ensembles would be equivalent.
This is indeed motivated by the analytical predictions for critical interfaces at depinning 
showing that the displacement field can be written as $u = \sqrt{\epsilon} u_0 + \epsilon u_1$ 
where $u_0$ is a Gaussian random variable of order ${\cal O}(1)$ while $u_1$ is a random variable with non-Gaussian
fluctuations, also of order ${\cal O}(1)$~\cite{width_frg}. 
Since $h(x)$ and ${\tilde h(x)}$ describe a self-affine interface with the same roughness exponent $\zeta$
we must have, in particular, 
\begin{eqnarray}
 \overline{h_m} &=& a L^{\zeta},\; \langle {\tilde h_m} \rangle = {\tilde a} L^{\zeta} \\
 \sqrt{\overline{(h_m-\overline{h_m})^2}}&=&  b L^{\zeta},\; 
 \sqrt{\langle (\tilde h_m - \langle \tilde h_m\rangle)^2 \rangle} = {\tilde b} L^{\zeta}.
\end{eqnarray}
For the particular case {\tt RPG} the parameters of the Gaussian signal ${\tilde a}$ and ${\tilde b}$ corresponding 
to~(\ref{exp_Fourier}) can be obtained analytically, since in this case the MRH distribution is Airy distributed, since 
$\zeta_{\tt RPG}=1/2$. 
We get ${\tilde a}_{\tt RPG} = \sqrt{\pi/8}$, ${\tilde b}_{\tt RPG}=\sqrt{5/12-\pi/8}$. For the other cases, 
the Gaussian parameters can be obtained numerically for a finite number of modes.  
In order to avoid mixing size effects present in both, the disordered interfaces and  
Gaussian signals through the number of modes, we have evaluated ${\tilde a}$ and ${\tilde b}$ for 
a very large number of modes. We find indeed that the value ${\tilde a}/{\tilde b}$ 
converges faster to its assymptotic value for larger roughness 
exponents. We have thus fixed {$L=4096$ modes for the Gaussian signals, 
assuring that the values ${\tilde a}$ and ${\tilde b}$ 
are almost numerically converged for the $\zeta=2/3$ case (and thus for all higher exponents).
By making the hypothesis~(\ref{eq:hypothesis}) we get
\begin{eqnarray}
 \overline{h_m} &=& \chi \langle {\tilde h_m} \rangle \Rightarrow \chi = \frac{a}{{\tilde a}}\\
 \overline{(h_m-\overline{h_m})^2} &=& \chi^2 \langle (\tilde h_m - \langle \tilde h_m\rangle)^2 \rangle 
\Rightarrow \chi = \frac{b}{{\tilde b}}  \;,
\end{eqnarray}
implying that $b/{\tilde b}=a/{\tilde a}$ if the hypothesis is true. To quantify the possible differences 
we can thus define the ratio 
\begin{equation}
\Delta = \frac{(a/{\tilde a})-(b/{\tilde b})}{\min \left( a/{\tilde a}, b/{\tilde b} \right)} \;,
\end{equation}
for each case $k$. In Fig.~\ref{fig_ratio} we show the evolution of $\Delta_k$ with the system size $L$ for all the 
cases, {\tt RMG}, {\tt RPG}, {\tt RMD}, and {\tt RPD}. As discussed above finite size effects come from $a$ and $b$ and not 
from ${\tilde a}$ and ${\tilde b}$.
We can 
see that $\Delta_k<0$ for the random-periodic cases ($k={\tt RPG, RPD}$) 
and the ground-state of the random-manifold ($k=\texttt{RMG}$),
while $\Delta_k>0$ for the random-manifold 
at depinning ($k={\tt RMD}$). 
{It indicates that the variance of the average-rescaled MRH distribution for the disordered interface 
data $k=\texttt{RPG}, \texttt{RMG}, \texttt{RPD}$
is larger than the variance of the corresponding average-rescaled MRH distribution for the Gaussian data,
and the opposite is true for the \texttt{RMD} case.}
{Focusing on the size-dependence, we see that $\Delta_{\tt RMD}$ quickly saturate 
to values of order of $0.03-0.04$, 
while both $|\Delta_{\tt RMG}|$ and $|\Delta_{\tt RPD}|$ slightly increase 
and should converge as well to a non-zero value for large systems $L\to \infty$. 
On the other hand 
$|\Delta_{\tt RPG}|$ slowly decreases with $L$ and show no tendency to saturation towards a finite value.
} 
Interestingly, as we show in the inset of Fig.~\ref{fig_ratio} we have that for the {\tt RPG} case 
$|\Delta_{\tt RPG}|\approx L^{-\theta_{\tt RPG}}$, with $\theta_{\tt RPG} \approx 0.5$. We find 
that this power law behaviour is not proper to the disordered interface however: the corresponding 
Gaussian signal also follows a very close power law as a function of the number of modes,
as its MRH distribution approaches the assymptotic Airy distribution.
\begin{figure}
\centering
\includegraphics[width=\linewidth]{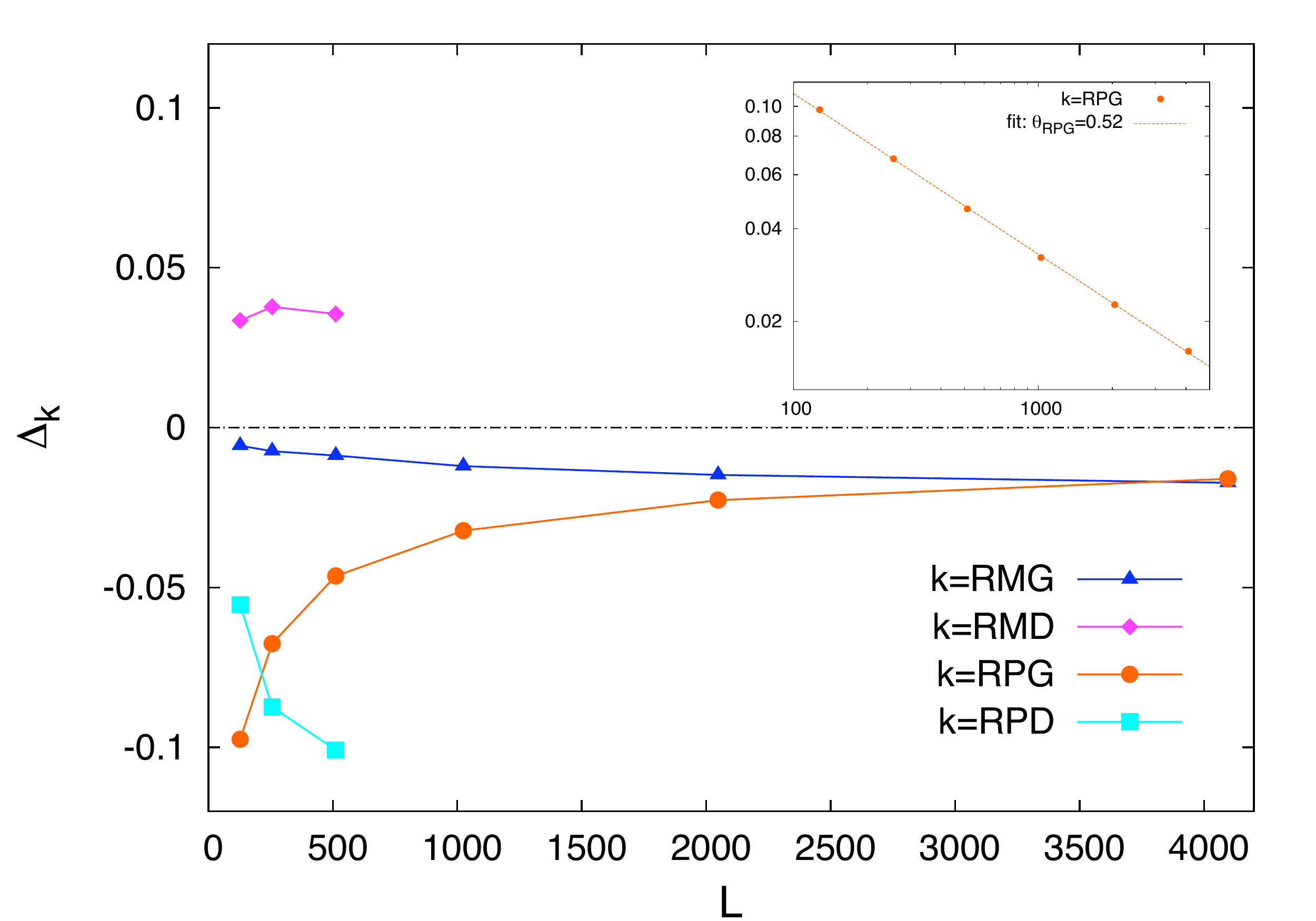}
\caption{Values of ratios $\Delta_k$ for each case $k={\tt RMG, RMD, RPG, RPD}$ plotted
with increasing size $L$ for the disordered interface data. The inset presents the power law
behavior of $|\Delta_{\texttt{RPG}}|$ towards zero, in log-log scale.}
\label{fig_ratio}
\end{figure}
To better quantify the non-Gaussian corrections we have performed a two sample 
Kolmogorov-Smirnov statistical test of equivalence between the MRH of sampled ground-states 
and their Gaussian approximations for the {\tt RMG} and {\tt RPG} cases. 
We have used the data sampled at the largest system sizes $L=2048,4096$ for both the disordered 
interfaces and Gaussian signals in order to reduce at maximum the possible 
finite-size effects in both processes. For the {\tt RMG} we find a statistically 
significant difference between the distributions, with a probability less than 
$p < 10^{-8}$ that the distributions are the same (null hypothesis). 
For {\tt RPG} however, we find a probability or significance of order $p=0.5$ for $L=2048$, and 
$p=0.8$ for $L=4096$. It is thus very probable that the two distributions are the same one. 
Since the Gaussian case tends to the Airy distribution as $L \to \infty$ we 
conclude that the MRH statistics in the Random-Periodic class 
is in the same class as the one of periodic normal random-walks with $\zeta=1/2$. 
In this respect we note that this extends the study made in~\cite{schehr_airy} 
made for a family of thermally equilibrated Solid-on-Solid models without disorder, 
to a disordered system at $T=0$.
\subsection{Boundary conditions}\label{boundary}

The dependence on boundary conditions is a relevant issue. On the one hand 
it is experimentally usually difficult to realize prescribed boundary conditions 
for self-affine interfaces. In this respect a theory describing 
``window boundary conditions'' have been recently proposed in which
the width statistics of segments become universal and independent of 
boundary conditions~\cite{raoul_width}. On the other hand, boundary 
conditions usually add technical difficulties to 
the analytical approaches. Numerical studies with 
different boundary conditions 
are thus interesting, as they allow us to know the sensitivity to boundary 
conditions of different observables. Here we will address 
their effects on the MRH statistics for particular cases.

So far we have only analyzed self-affine periodic signals $h(0)=h(L)$ with a given period $L$, either 
originated from independent Fourier modes or from the interplay between 
local elasticity,  disorder and a driving force, which all 
have stationary spatial correlation functions. 
This means for instance that for any two observables ${\cal O}_1(x)$ and ${\cal O}_2(x)$ 
the Gaussian signals considered satisfy 
$\langle {\cal O}_1(x){\cal O}_2(x') \rangle \equiv \langle {\cal O}_1(0){\cal O}_2(x'-x) \rangle$, 
while the considered critical configurations at depinning and ground-state configurations 
satisfy $\overline{{\cal O}_1(x){\cal O}_2(x')} \equiv \overline{{\cal O}_1(0){\cal O}_2(x'-x)}$. 
In particular, this implies that the MRH can occur, with uniform probability, at any position $x$.

Non-stationary spatial correlation functions can be easily generated using 
independent Fourier modes, while keeping the self-affine geometry, by applying different 
boundary conditions. For a self-affine signal $h(x)$ with roughness exponent $\zeta$ we can 
for instance impose fixed-ends $h(0)=h(L)=0$ by constructing a pure sine series 
${\tilde h}(x) = \sum_{n=1} c_n \sin(n \pi x /L)$ with normally distributed uncorrelated 
amplitudes $c_n=N[0,(2/\pi n) (L/\pi n)^{\zeta}]$~\cite{wernerbook}. 
It is easy to see that these signals have in general
non-stationary spatial correlation functions. Consider, for instance, the observables ${\cal O}_{1}(x) = {\cal O}_{2}(x)=h(x)-h(0)$: the correlation function
$\langle {\cal O}_1(x){\cal O}_2(x') \rangle$ vanishes for $x=0$ (or $x'=0$) and $x=L$ (or $x'=L$). In particular, it is clear that for fixed-ends the location of the 
MRH of each signal is not uniformly distributed along $L$: instead its pdf is peaked around $x=L/2$. The MRH distribution of these kind of non-stationary 
signals is not known exactly in general.

Ground-state configurations of elastic interfaces in random media can also have non-stationary spatial correlations 
functions in presence of certain constraints. Indeed we have seen that an optimal ground-state path with $h(0)=h(L)$ [note that we do not consider in this paper ``tilted'' 
signals with $h(0)\neq h(L)$] can be generated in two ways. 
The one we have analyzed so far comes from a minimization over all possible paths in the disordered substrate 
with $h(0)=h(L)$, and periodic boundary conditions in the displacements direction. This yields stationary 
ground-state configurations. On the other hand, if we do not seek the minimum among {\it all} possible paths 
such that $h(0)=h(L)$, but fix the extremity $h(0)=h(L)$ 
to a particular value we obtain ground-state configurations with non-stationary spatial correlation functions. In the 
two cases however,  the self-affine structure is preserved, as it only depends on the variance of Fourier modes 
amplitudes rather than in its
phases. Finally, let us note that the critical configurations at depinning we have analyzed have, 
by construction, stationary spatial correlation functions.

We will address here the effects of boundary conditions in the MRH distribution. To this purpose 
we focus on the well known directed polymer in a random medium, equivalent to the {\tt RMG} case for our 
disordered elastic manifold. To illustrate the effects in this case 
we generate two particular types of 
boundary conditions by using the two methods 
outlined above to obtain both stationary and non-stationary ground-states 
with $\zeta=2/3$. One may write the height field as a fully periodic series, containing
both cosines and sines as in Eq. (\ref{exp_Fourier}) in the stationary case and as a sine series in the non-stationary case.   
We note that these boundaries conditions would give equivalent results
for $\zeta=1/2$~\cite{wernerbook}, which is the case of the {\tt RPG} disordered interface.

In Fig.~\ref{fig_boundary} we plot the difference of the cumulative MRH distributions in average 
scaling for the two cases, $\Delta F_{\texttt{RMG'}}(x)=F_{\texttt{RMG}}(x)-F_{\texttt{RMG'}}(x)$,
where {\tt RMG} and {\tt RMG'} denote the stationary and non-stationary ground-states 
respectively of the same size $L$, for $L=128,256,512$. 
We observe that for all sizes the difference is appreciable, of the same order than the 
difference between the {\tt RMG} case and its Gaussian approximation, and it does 
not have any appreciable decrease with increasing $L$. We thus conclude that the 
MRH distribution of RM ground-states is sensitive to the boundary conditions.

{In order to understand the origin of the sensitivity to boundary conditions 
we have also compared the MRH statistics of Gaussian interfaces described 
by the sine series (defined above) and by the full periodic series of Eq.~(\ref{exp_Fourier}).} 
A Kolmogorov-Smirnov statistical test 
over $10^6$ averaged-scaled numerical Gaussian realizations shows that the MRH distribution for the 
sine series has, for $\zeta=1/2$, a probability $p\sim 0.9$ of being the same than the MRH distribution of the 
full-periodic series, while for $\zeta=2/3$ the same probability is $p \sim 10^{-11}$. 
While for $\zeta=1/2$ these results can be simply related to the fact that the signals are Markovian 
({and thus quickly loose ``memory'' of the border}), 
for $\zeta=2/3$ they show that the dependence on boundary conditions {is indeed closely 
related to the anomalous geometry, rather than to its particular microscopic origin or 
to the presence of non-Gaussian corrections. 
This extends the results obtained for the width distribution 
of Gaussian signals~\cite{raoul_width} to the case of the MRH.}
\begin{figure}[h]
\includegraphics[width=\linewidth]{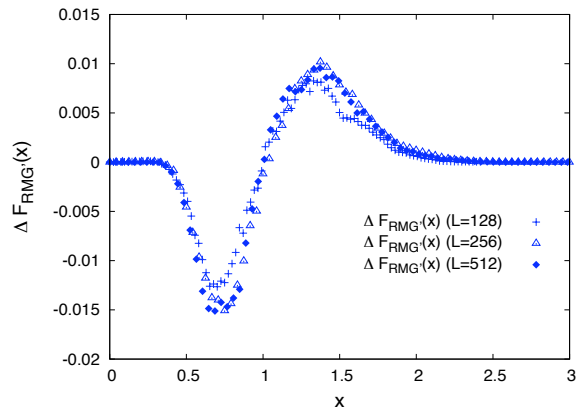}
\caption{Differences $\Delta F_{\texttt{RMG'}}(x)=F_{\texttt{RMG}}(x)-F_{\texttt{RMG'}}(x)$ 
of cumulative distributions between the two ways of
selecting the optimal path in a disordered medium, for lengths $L=128,256,512$. 
The cumulative distributions are computed with average scaling, \textit{i.e.} with $x=h_m/\overline{h_m}$.}
\label{fig_boundary}
\end{figure}

\section{Conclusions}

We have studied the maximal relative height statistics
of self-affine one-dimensional elastic interfaces in a random 
potential. We have analyzed, by exact numerical sampling methods, 
the ground state configuration and the critical configuration at 
depinning, both for the experimentally relevant 
random-manifold and random-periodic universality classes. 
We have also analyzed the MRH distribution
of self-affine signals generated by independent Gaussian Fourier modes and 
obtained an exact analytical description for the right tails using Pickands' theorem. 
We found that in general, the independent Gaussian modes model 
provides a good approximation for predicting the MRH statistics, as it was 
previously found for the width distribution at depinning. This result 
is a priori not obvious as the MRH is an extreme observable (occurring at 
a single point in space for each realization) while the width is 
a spatially averaged one.

The comparison of the different MRH distributions with their approximation using 
Gaussian independent Fourier modes shows small differences that quickly saturate with system size 
in the case of depinning, in agreement with the results obtained for the width distribution 
of these configurations~\cite{width_rosso,width_frg}: this confirms 
the predictions of non-Gaussian corrections at depinning. It would be very interesting to characterize
more quantitatively these non-Gaussian corrections that we have observed for the MRH at the depinning 
threshold using the tools of Functional
Renormalization Group~\cite{review_frg}. A preliminary analysis of this question indicates that this
is not a simple extension of the previous works on the distribution of the width~\cite{width_rosso, width_frg}: indeed the extension of the perturbation
theory proposed in Refs.~\cite{width_rosso, width_frg} to the computation of the MRH distribution involves, at lowest order, a
constrained propagator (with an absorbing boundary at $h=0$) for which the Wick's theorem does not hold.

On the other hand, for ground-state configurations we also find small differences but 
they display a slower decrease with system size. 
For the random-manifold ground-states a finite difference 
is found in the large size limit, implying non-Gaussian corrections. 
This is qualitatively consistent with analytical results 
obtained for other physical observables for the directed polymer in a random medium. For instance, 
the energy of the optimal polymer which is described by one of the Tracy-Widom
distributions (associated to the Gaussian Unitary Ensemble or to the Gaussian Orthogonal Ensemble of random matrices depending on the
geometry of the problem), which thus shows strong deviations
from Gaussian fluctuations~\cite{dprm}.
The random-periodic ground 
states on the other hand have a MRH distribution that is indistinguishable 
from the Airy distribution for the largest system sizes, as follows from 
studying its moments and by a Kolmogorov-Smirnov statistical test. This further 
confirms the universality of the Airy distribution for periodic signals 
with $\zeta=1/2$, as shown for a family of non-disordered thermally equilibrated 
one-dimensional solid-on-solid models~\cite{schehr_airy}. Our results are also consistent
with previous results obtained for the related Solid-on-Solid model on a disordered substrate in $1+1$ dimensions 
(which belongs to the same universality class as RPG)~\cite{solidonsolid}. For this model 
with one free end, the ground state
configuration can be constructed iteratively so that the height field behaves, 
on large scale, as a random walk with $\zeta = 1/2$. Notice however
that at variance with Ref. \cite{solidonsolid} we have considered 
here periodic boundary conditions.

Finally, we have shown that MRH distributions are in general 
sensitive to boundary conditions. 
This might be important 
for experiments, where particular boundary conditions are
difficult to realize or are not precisely known. 
In this respect it would be interesting
to perform an analysis of the MRH statistics using the 
``window boundary conditions'' or ``segment statistics'' 
proposed and applied to the width statistics in Ref.~\cite{raoul_width}.

Our results show that the MRH statistics provide a valuable tool 
to study experimental images of self-affine interfaces, such as 
magnetic domain walls or contact lines in partial wetting. 
In particular, it allows to infer information about the 
mechanisms behind the universal self-affine geometry, 
such as the disorder-induced coupling between Fourier modes, 
or more generally the one induced by non-linear interaction 
terms. It also allows to infer information about 
boundary conditions. Our results might be used as a guide for 
analytical approaches predicting the statistics of extreme 
geometrical observables.

\begin{acknowledgments}
{We acknowledge Satya N. Majumdar and Alberto Rosso for useful discussions.} 
This work was supported by the France-Argentina MINCYT-ECOS A08E03. 
A.B.K acknowledges the hospitality at LPT-Orsay, J. R. and G. S acknowledges the hospitality at the Centro Atomico in Bariloche.   
S.B. and A.B.K acknowledge support from CNEA, CONICET under Grant No. PIP11220090100051, and
ANPCYT under Grant No. PICT2007886. 

\end{acknowledgments}

\newpage

\appendix

\section{Pickands' theorem and the right tail of the distribution of the MRH for a Gaussian path}\label{appendix_A}

Let $X(t)$, $t \in [0,T]$ be a continuous centered Gaussian process with covariance function $r(t) = \langle X(s) X(s+t)\rangle$ which
satisfies 
\begin{enumerate}
\item{$r(t) \leq 1$, for $t \in [0,T]$ \;,}
\item{$r(t) = 1 - C |t|^\mu + o(|t|^\mu)$ as $t \to 0$ \; \;,}
\end{enumerate}
where $T > 0$, $\mu \in (0,2]$ and $C > 0$ are constants. Let us define 
\begin{eqnarray}
X_{\max} = \max_{0 \leq t \leq T} X(t) \;.
\end{eqnarray}
Pickands' results concerns the right tail of the distribution of $X_{\max}$, ${\rm Pr}(X_{\max} \leq \xi)$~\cite{pickands}:
\begin{eqnarray}\label{th_pickands}
\lim_{\xi \to \infty} \frac{{\rm Pr}(X_{\max} \geq \xi)}{\xi^{2/\mu} \Phi(\xi)} = T C^{1/\mu} H_{\mu} \;, \\ \nonumber
\end{eqnarray}
where $\Phi(\xi) = \frac{1}{\sqrt{2 \pi}} \int_\xi^\infty e^{-u^2/2}$ and $H_{\mu} > 0$, the so-called Pickands' constant, is given by
\begin{eqnarray}\label{pickands_constant}
&&H_{\mu} = \lim_{T \to \infty} T^{-1} \int_0^\infty e^{s} {\rm Pr}\left(\max_{0\leq t \leq T} Y(t) > s\right) \, \rmd s \;, \nonumber \\
&& Y(t) = \sqrt{2} B_{\mu/2}(t) - t^\mu
\end{eqnarray}
where $B_{\mu/2}(t)$ is the fractional Brownian motion with Hurst exponent $\mu/2$, {\it i.e.} the Gaussian process characterized by the two-point correlation function:
\begin{eqnarray}
\langle B_{\mu/2}(t) B_{\mu/2}(s)\rangle = \frac{1}{2}\left(t^\mu + s^\mu - |t-s|^\mu\right) \;.
\end{eqnarray} 
No explicit formula exist for $H_\mu$ except for $\mu = 1$ and $\mu = 2$, for which $H_{1} = 1$ and $H_2 = 1/\sqrt{\pi}$. The result above (\ref{th_pickands}) means that the pdf of $X_{\max}$ behaves, for large argument as
\begin{eqnarray}\label{expr_pickands}
\frac{d}{d\xi} {{\rm Pr}(X_{\max} \leq \xi)} \sim T C^{1/\mu} \frac{H_\mu}{\sqrt{2 \pi}} \xi^{2/\mu} e^{-\frac{\xi^2}{2}} \;.
\end{eqnarray}

Here we will apply this result (\ref{expr_pickands}) to derive the asymptotic behavior of $h_{m} = \max_{0 \leq x \leq L} h(x)$ where $h(x)$ is given in Eq. (\ref{exp_Fourier}) and distributed according a Gaussian probability measure as in Eq. (\ref{Gauss}). We thus first compute the two-point correlation function as
\begin{equation}\label{correl_1}
 \langle h(x) h(x')\rangle = \frac{2 L^{\alpha-1}}{(2 \pi)^\alpha} \sum_{n=1}^{\infty} \frac{1}{n^{\alpha}} \cos{\left(\frac{2\pi n}{L}(x-x') \right)},
\end{equation}
which is obviously stationary. One has in particular ${\rm w}^2_L = \langle h(x) h(x)\rangle = \frac{2 L^{\alpha-1}}{(2 \pi)^\alpha} {\bm \zeta}(\alpha)$, independently of $x$. If one defines
\begin{eqnarray}
\tilde h(x) = \frac{h(x)}{{\rm w}_L} \;,
\end{eqnarray}
then the two-point correlator is also stationary $r(x) = \langle \tilde h(y) \tilde h(y+x)\rangle$ and periodic $r(0) = r(L)$. In addition one has from Eq.~(\ref{correl_1}) that it satisfies $r(x) \leq 1$, $\forall x \in [0,L]$. To apply Pickands' theorem we need to analyze the small $x$ behavior of $r(x)$. To this purpose, it is useful to write $r(x)$ as
\begin{equation}\label{correl_2}
 r(x) = \frac{1}{2\zeta(\alpha)} \left[ {\rm Li} \left(\alpha,e^{2 i \pi x/L}\right) + {\rm Li}\left(\alpha,e^{-2 i \pi x/L}\right)  \right] \;,
\end{equation}
where ${\rm Li}(\alpha,x)$ is the polylogarithm function \cite{abramowitz}. In particular, for $\alpha = 2n$ one has \cite{abramowitz}
\begin{eqnarray}
r(x) &=& (-1)^{n+1} \frac{(2 \pi)^{2n}}{2(2n)! {\bm \zeta}(2n)} B_{2n}(x) \nonumber \\
 &=& \frac{B_{2n(x)}}{B_{2n}(0)} \;,
\end{eqnarray}
where $B_{2n}(x)$ is the Bernoulli polynomial of degree $2n$ and $B_{2n}(0) = {\rm b}_{2n}$ is a Bernoulli number, which satisfies
${\rm b}_{2n} = 2 (-1)^{n-1} {\bm \zeta}(2n) (2n)!/(2\pi)^{2n}$. Hence one has for instance, for $\alpha=2$
\begin{eqnarray}
r(x) = 1 - 6x + 6x^2 \;,
\end{eqnarray}
so that, in that case, $\mu = 1$ while for $\alpha=4$ one has
\begin{eqnarray}
 r(x) = 1 - 30 x^2 + 60 x^3 - 30 x^4 \;,
\end{eqnarray}
so that $\mu =2$ in that case. The small argument behavior of $r(x)$ can be obtained for any $\alpha$ from
a careful asymptotic analysis of Eq. (\ref{correl_2}) which yields
\begin{eqnarray}
 r(x) \sim 
\begin{cases}
&1 - 2\pi \frac{|\Gamma(1-\alpha) \sin{\frac{\alpha \pi}{2}}|}{{\bm \zeta}(\alpha)} x^{\alpha-1} \;, \; 1 < \alpha < 3 \:, \\
&1 + \frac{2 \pi^2}{{\bm \zeta}(3)}x^2 \log{x} \;, \; \alpha =3 \;, \\
& 1 - 2\pi^2 \frac{{\bm \zeta}(\alpha-2)}{{\bm \zeta}(\alpha)} x^2 \;, \; \alpha > 3 \;.
\end{cases}
\end{eqnarray}
It is then straightforward to use Pickands' theorem (\ref{expr_pickands}) to obtain the results given in the text in Eqs (\ref{gauss_large}, \ref{delta}, \ref{ampli_D}, \ref{ampli_E}). Note that in the general case where $1 < \alpha < 3$, one obtains the amplitude $D$ as
\begin{eqnarray}\label{gen_alpha}
D &=& \frac{H_{\alpha-1}}{\sqrt{2 \pi}} \left(\frac{2 \pi}{\zeta(\alpha)} \left|\Gamma(1-\alpha) \sin{\left(\frac{\alpha \pi}{2}\right)}   \right|  \right)^{1/(\alpha-1)} \nonumber \\ 
&\times& \left(\frac{(2 \pi)^\alpha}{2 \zeta(\alpha)} \right)^{\frac{\alpha+1}{2(\alpha -1)}} 
\end{eqnarray}
where $H_{\alpha-1}$ is the Pickands' constant given in Eq. (\ref{pickands_constant}).

\bibliographystyle{apsrev4-1.bst}


\begin{thebibliography}{100}

\bibitem{bouchaud_mezard}
J.-P. Bouchaud and, M. M\'ezard, J. Phys. A {\bf 30}, 7997 (1997).

\bibitem{bouchaud_biroli_review}
G. Biroli, J.-P. Bouchaud, and M. Potters, J. Stat. Mech. P07019 (2007).

\bibitem{RCPS}
S. Raychaudhuri, M. Cranston, C. Przybyla, and Y. Shapir, Phys. Rev. Lett. {\bf 87}, 136101 (2001).

\bibitem{satya_airyshort}
S. N. Majumdar and A. Comtet, Phys. Rev. Lett. {\bf 92}, 225501 (2004).

\bibitem{satya_airylong}
S. N. Majumdar and A. Comtet, J. Stat. Phys. {\bf 119}, 777 (2005).

\bibitem{schehr_airy}
G. Schehr and S. N. Majumdar, Phys. Rev. E {\bf 73}, 056103 (2006).

\bibitem{GMOR}
G. Gy\"orgyi, N. R. Moloney, K. Ozog{\'a}ny, and Z. R{\'a}cz, Phys. Rev. E {\bf 75}, 021123 (2007).

\bibitem{RS09}
J. Rambeau and G. Schehr, J. Stat. Mech. P09004 (2009).

\bibitem{RS10}
J. Rambeau and G. Schehr, Europhys. Lett. {\bf 91}, 60006 (2010); Phys. Rev. E {\bf 83}, 061146 (2011).

\bibitem{barabasi_stanley}
A. L. Barab{\'a}si and H. E. Stanley, {\it "Fractal concepts in surface growth"} (Cambridge University Press, 1995). 

\bibitem{width_gaussian}
T. Antal, M. Droz, G. Gy\"orgyi, and Z. R{\'a}cz, Phys. Rev. Lett. {\bf 87}, 240601 (2001); Phys. Rev. E {\bf 65}, 046140 (2002). 

\bibitem{width_rosso}
A. Rosso, W. Krauth, P. Le Doussal, J. Vannimenus, and K.~J. Wiese, Phys. Rev. E {\bf 68}, 036128 (2003).

\bibitem{width_frg}
P. Le Doussal and K. J. Wiese, Phys. Rev. E {\bf 68}, 046118 (2003). 

\bibitem{review_frg}
P. Le Doussal, Ann. Phys. {\bf 325}, 49 (2010);
K. J. Wiese and P. Le Doussal, Markov Processes Relat. Fields {\bf 13}, 777 (2007).

\bibitem{moulinet_width}
S. Moulinet, A. Rosso, W. Krauth, and E. Rolley, Phys. Rev. E {\bf 69}, 035103(R) (2004). 

\bibitem{raoul_width}
R. Santachiara, A. Rosso, and W. Krauth, J. Stat. Mech. P02009 (2007).

\bibitem{majumdar_review}
S. N. Majumdar, Current Science {\bf 89}, 2076 (2005).

\bibitem{domain_walls}
S.~Lemerle, J.~Ferr\'e, C.~Chappert, V.~Mathet, T.~Giamarchi, and P. Le Doussal, Phys. Rev. Lett. {\bf 80}, 849 (1998);
M.~Bauer, A.~Mougin, J.~P.~Jamet, V.~Repain, J.~Ferr\'e, S.~L.~Stamps, H.~Bernas, and C. Chappert, Phys. Rev. Lett. {\bf 94}, 207211 (2005);
M.~Yamanouchi, D.~Chiba, F.~Matsukura, T.~Dietl, and H.~Ohno, Phys. Rev. Lett. {\bf 96}, 096601 (2006).

\bibitem{wetting}
P. Le Doussal, K. J. Wiese, S. Moulinet, and E. Rolley, Europhys. Lett. {\bf 87}, 56001 (2009).

\bibitem{fracture}
M. Alava, P. K. V. V. Nukalaz, and S. Zapperi, Adv. Phys. {\bf 55}, 349 (2006);
L. Ponson, D. Bonamy, and E. Bouchaud, Phys. Rev. Lett. {\bf 96}, 35506 (2006); 
D. Bonamy, S. Santucci, and L. Ponson, Phys. Rev. Lett. {\bf 101}, 045501 (2008).

\bibitem{cdw}
T. Nattermann and S. Brazovskii, Adv. Phys. {\bf 53}, 177 (2004). 

\bibitem{blatter}
G. Blatter, M. V. Feigelman, V. B. Geshkenbein, A. I. Larkin, and V. M. Vinokur, Rev. Mod. Phys. {\bf 66}, 1125 (1994). 

\bibitem{giamarchi}
T. Giamarchi, S. Bhattacharya, in {\it High Magnetic Fields: Applications in Condensed Matter Physics and Spectroscopy}, Ed. C. Berthier et al. (Springer-Verlag, Berlin, 2002) p. 314, cond-mat/0111052.

\bibitem{phasediagram}
A. B. Kolton, A. Rosso, T. Giamarchi, and W. Krauth Phys. Rev. B {\bf 79}, 184207 (2009);
Phys. Rev. Lett. {\bf 97}, 057001 (2006). 

\bibitem{kardar}
M. Kardar, Nucl. Phys. B {\bf 290}, 582 (1987);
D.~A. Huse, C.~L. Henley, and D.~S. Fisher, Phys. Rev. Lett. {\bf 55}, 2924 (1985);
M. Kardar and Y.-C. Zhang, Phys. Rev. Lett. {\bf 58}, 2087 (1987);
T. Halpin-Healy and Y.-C. Zhang, Phys. Rep. {\bf 254}, 215 (1995).

\bibitem{rmtorp}
S. Bustingorry, A. B. Kolton, and T. Giamarchi, Phys. Rev. B {\bf 82}, 094202 (2010).

\bibitem{solidonsolid}
H. Rieger and U. Blasum, Phys. Rev. B {\bf 55}, R7394 (1997).

\bibitem{rosso_1.25}
A. Rosso, A. K. Hartmann, and W. Krauth, Phys. Rev. E {\bf 67}, 021602 (2003).

\bibitem{rmtorp2}
S. Bustingorry and A. B. Kolton, Papers in Physics {\bf 2}, 020008 (2010).

\bibitem{chauve}
P. Chauve, T. Giamarchi, and P. Le Doussal, Phys. Rev. B {\bf 62}, 6241 (2000).


\bibitem{pickands}
J. Pickands III, Trans. Amer. Math. Soc. {\bf 145}, 75 (1969).

\bibitem{abramowitz}
M. Abramowitz and I.~A. Stegun in {\it Handbook of Mathematical Functions} (Dover, New York, 1973).

\bibitem{takacs_invert} 
L.~Takacs, J. Appl. Prob. {\bf 32}, 375 (1995).

\bibitem{janson_louchard}
S. Janson and G. Louchard, Elec. Journ. Prob. {\bf 12}, 1600 (2007).



\bibitem{hilhorst_sylvester}
H.~J.~Hilhorst, P. Calka, and G. Schehr, J. Stat. Mech. P10010, (2008).

\bibitem{rosso_algorithm}
A. Rosso and W. Krauth, Phys. Rev. Lett. {\bf 87}, 187002 (2001); 
A. Rosso and W. Krauth, Phys. Rev. B {\bf 65}, 012202 (2002).

\bibitem{wernerbook}
W. Krauth, {\it Statistical Mechanics: Algorithms and Computations} (Oxford University Press, Oxford, 2006) (See
www.phys.ens.fr/doc/SMAC).

\bibitem{dprm}
M. Pr\"ahofer and H. Spohn, Phys. Rev. Lett. {\bf 84}, 4882 (2000); J. Stat. Phys. {\bf 108}, 1071 (2002);
K.~Johansson, Comm.~Math.~Phys. {\bf 209}, 437 (2000);
P. Calabrese, P. Le Doussal, and A. Rosso, Europhys. Lett. {\bf 90}, 20002 (2010);
V. Dotsenko, Europhys. Lett. {\bf 90}, 20003 (2010);
P. Calabrese and P. Le Doussal, Phys. Rev. Lett. {\bf 106}, 250603 (2011);
P. J. Forrester, S. N. Majumdar, and G. Schehr, Nucl. Phys. B {\bf 844}, 500 (2011).





\end{thebibliography}

\end{document}